\documentclass[12pt]{article}
\usepackage{a4,amsmath,float,graphicx}
\usepackage[square,comma,sort&compress]{natbib}
\floatstyle{plain}
\restylefloat{table}
\restylefloat{figure}
\date{}
\begin{document}
\title{Singlet and Triplet Differential Cross Sections for 
$pp\to pp\pi^\circ$}
\maketitle
{\centering{P. N. Deepak and G. Ramachandran\footnote[1]{Present Address: Indian Institute
of Astrophysics, Koramangala, Bangalore-560 034, India}\\[.2cm]
{\it Department of Studies in Physics, University of 
Mysore, Mysore-570 006, India}}}

\vspace*{2cm}

\begin{abstract}
The singlet and triplet differential cross sections for $pp\to pp\pi^\circ$
have been estimated for the first time at 325, 350, 375 and 400 MeV using 
the results of the recent experimental measurements 
[Phys. Rev. C \textbf{63}, 064002 (2001)] of Meyer et al.\newline

\vspace*{.5cm}

\end{abstract}

\vfill\newpage

The study of the reaction ${\vec p}{\vec p}\to pp\pi^\circ$ has 
excited considerable 
experimental interest \cite{mey1,mey2,thorn,mey3}, since the 
transition is mainly to the final $Ss$ state and is completely
spin-dependent at
threshold energies up to 300 MeV and the large values of momentum 
transfer involved probes the interaction at very short distances.  As the 
energy increases, the transitions to $Ps$ and $Pp$ states are 
expected to contribute \cite{mey1,mey2} below
400 MeV although pionic $d$-wave effects have been reported 
\cite{zlom} even at a beam energy 
of 310 MeV.  Taking, therefore, transitions to $Sd$ and $Ds$ 
states also into account, an analysis of the measurements \cite{mey3} 
on a complete set of polarization observables have been presented recently 
for energies below 400 MeV.  Attention has also been drawn \cite{mey4} 
to the large longitudinal analyzing power characterizing this 
reaction.  It is of interest to note \cite{gr1} that 
the difference $^{3}d^{2}\sigma_{+1}-^{3}d^{2}\sigma_{-1}$ is indeed
proportional to the longitudinal analysing power, where $^{2s_i+1}d^{2}
\sigma_{m}$ denotes the differential cross section for the process 
$pp\to pp\pi^\circ$ as a function of five independent kinematic variables 
characterizing the three-body final state when the
reaction is initiated in the spin state $\vert s_i,m\rangle,\ m=-s_i,\ldots,+s_i$
and $s_i=0,1$ corresponding to singlet and triplet states.   Measurements have 
also been reported \cite{bilger} at energies
up to 425 MeV which were analysed in terms of the above mentioned 
partial waves, where evidence for $Ds$ state was seen through the presence of 
a $\cos^{4}\theta$ term even at 310 MeV.  In this context, it is of 
interest to note that analysis of the total cross section $\sigma$ into its 
singlet and triplet components $^{2s_i+1}\sigma_{m}$ have been attempted in 
several ${\vec p}{\vec p}\to pp\pi^\circ$ measurements \cite{mey1,mey2,thorn} 
using the theoretical results of Bilenky and Ryndin \cite{br}, which however 
are not applicable at the differential level.  On the other hand, it may also 
be noted that the irreducible tensor approach \cite{gr2} to pion production in 
$NN$ collisions leads to elegant formulas for the differential cross 
sections in terms of irreducible tensor amplitudes $M^\lambda_{\mu}(s_{f},s_{i})$ 
of rank $\lambda=|s_i-s_f|$ to $(s_i + s_f)$, where $s_i, s_f$ denote the 
initial and final channel spins.  This approach, moreover, is not limited by 
the number of final partial waves.
The purpose of this Brief Report is therefore, to present estimates,
using \cite{gr1,gr2} for the singlet and triplet cross sections at the differential level 
itself based on the recently
reported measurements of Meyer et al \cite{mey3}.

We first of all note that the experimentally measured 
differential cross section for ${\vec p}{\vec p}\to pp\pi^\circ$ 
may be written, using the same notations as in \cite{mey3}, as
\begin{equation}
\label{relate}
{\frac{d\sigma(\theta_{p},\varphi_{p},
\theta_{q},\varphi_{q},\epsilon,{\vec P},
{\vec Q})}
{d\Omega_{p}d\Omega_{q}d\epsilon}}\equiv\sigma(\xi,{\vec P},{\vec Q})
=\textrm{Tr}\left( \textbf{M}\rho\textbf{M}^\dagger\right)
\end{equation}
where $\rho$
denotes the initial spin density matrix 
\begin{equation}
\label{rho}
\rho={\textstyle{\frac{1}{4}}}\left(1+\vec{\sigma}_{1}\cdot
{\vec P}\right)\left(1+\vec{\sigma}_{2}\cdot
{\vec Q}\right)
\end{equation}
and $\textbf{M}$ has the form \cite{gr2,gr1}
\begin{equation}
\textbf{M}=\sum_{s_i,s_f=0,1} \sum_{\lambda=|s_i - s_f|}^{s_i + s_f}
\left(S^\lambda(s_f,s_i)\cdot M^\lambda(s_f,s_i)\right),
\end{equation}
in terms of the irreducible tensor operators $S^\lambda _\mu(s_f,s_i)$
of rank $\lambda$, as defined in \cite{gr3}. The irreducible tensor
amplitudes $M^\lambda _\mu(s_f,s_i)$ may be expressed
following \cite{gr2} in terms of the partial wave
reaction amplitudes $\mathsf{M}^J_{l_{q}(l_{p}s_f)j;ls_i}(s,s_{12})$, 
employing the same notations for
the angular momenta and the Mandelstam variables as in
\cite{mey3}, through 
\begin{align}
\label{par-wave}
M^\lambda _\mu(s_f,s_i)=\sum_{l,l_{p},l_{q},L_{f},j,J}\ W(ls_{i}L_{f}&s_{f};J\lambda)
\ C(L_f l\lambda;\mu 0\mu)\ \mathsf{M}^J_{l_{q}(l_{p}s_f)j;ls_i}(s,s_{12})
\nonumber\\&\times
\left(Y_{l_p}(\theta_p,\varphi_p)\otimes
Y_{l_q}(\theta_q,\varphi_q)\right)^{L_f}_{\mu}.
\end{align}

We next express $\rho$ in the channel spin representation in the form
\begin{equation}
\label{rho-i-channel-spin}
\rho=\sum_{s_i,s_{i}'=0}^{1}\ \sum_{k=\vert s_i-s_{i}'\vert}^{(s_{i}+s_{i}')}
\left(I^{k}(s_{i},s_{i}')\cdot S^{k}(s_{i},s_{i}')\right),
\end{equation}
where
\begin{equation}
\label{I}
I^{k}_{q}(s_{i},s_{i}')={\textstyle{\frac{1}{2}}}[s_{i}']\ \sum_{k_{1}=0}^{1}
\ \sum_{k_{2}=0}^{1}(-1)^{k_{1}+k_{2}-k}[k_{1}][k_{2}]
\left\{\begin{matrix}
{\textstyle{\frac{1}{2}}} & {\textstyle{\frac{1}{2}}} & s_{i}\\[.2cm]
{\textstyle{\frac{1}{2}}} & {\textstyle{\frac{1}{2}}} & s_{i}'\\[.2cm]
k_{1} & k_{2} & k\\
\end{matrix}
\right\}
\ \left(P^{k_{1}}\otimes Q^{k_{2}}\right)^{k}_{q},
\end{equation}
with $P^0_0=Q^0_0=1$.

Using the known properties \cite{gr3} of the irreducible tensor 
operators $S^{k}_{q}$,
the differential cross section given by Eq. ({\ref{relate}}) may 
then be expressed in the form
\begin{equation}
\label{dcs-channel-spin}
\sigma(\xi,{\vec P},{\vec Q})=
\sum_{s_{i},s_{i}'=0}^{1}\ \sum_{k}
\left(I^{k}(s_{i},s_{i}')\cdot {\mathcal B}^{k}(s_{i},s_{i}')\right),
\end{equation} 
where the irreducible tensors 
\begin{align}
\label{bilinears}
{\mathcal B}^{k}_{q}(s_{i},s_{i}')=
[s_{i}]\sum_{s_{f}}&[s_{f}]^{2}\sum_{\lambda,\lambda'}(-1)^{\lambda}[\lambda]
[\lambda']
\ W(s_{i}'ks_{f}\lambda;s_{i}\lambda')
\nonumber\\&\times \left(M^{\lambda}(s_{f},s_{i})\otimes M^{\dagger^{\lambda'}}(s_{f},s_{i}')
\right)^{k}_{q}
\end{align}
are bilinear in terms of the irreducible tensor amplitudes.  They
can be explicitly evaluated using Eq. (\ref{par-wave}) for
transitions to $Ss,\ Ps,\ Pp,\ Sd$ and $Ds$ states.  

The unpolarized differential cross section $\sigma_{0}(\xi)$ for 
the process is then given by
\begin{equation}
\label{udcs}
\sigma_{0}(\xi)=\sum_{s_{i}=0,1}\ \sum_{m=-s_{i}}^{s_{i}}
{^{2s_{i}+1}}\sigma_{m}(\xi),
\end{equation}
where
\begin{subequations}
\label{ch-spin-cs}
\begin{align}
\label{ch-spin-1}
^{1}\sigma_{0}(\xi)&={\textstyle{\frac{1}{4}}}\ {\mathcal B}^{0}_{0}(0,0)\\
\label{ch-spin-2}
^{3}\sigma_{0}(\xi)&={\textstyle{\frac{1}{4}}}\left[
{\textstyle{\frac{1}{3}}}{\mathcal B}^{0}_{0}(1,1)-{\textstyle{\frac{\sqrt 2}{3}}}
{\mathcal B}^{2}_{0}(1,1)
\right]\\
\label{ch-spin-3}
^{3}\sigma_{\pm 1}(\xi)&={\textstyle{\frac{1}{4}}}\Bigl[
{\textstyle{\frac{1}{3}}}{\mathcal B}^{0}_{0}(1,1)\pm{\textstyle{\frac{1}{\sqrt 6}}}
{\mathcal B}^{1}_{0}(1,1)+{\textstyle{\frac{1}{3{\sqrt 2}}}}{\mathcal B}^{2}_{0}(1,1)
\Bigr].
\end{align}
\end{subequations}
Using standard Racah algebra, we also have explicitly
\begin{subequations}
\label{H}
\begin{align}
\label{b-singlet}
{\mathcal B}^{0}_{0}(0,0)&=A_{1}\cos^{2}\theta_{p}+A_{2}\sin^{2}\theta_{p}\\
{\mathcal B}^{0}_{0}(1,1)&=B_{0}+B_{1}(3\cos^{2}\theta_{q}-1)+B_{2}(3\cos^{2}\theta_{p}-1)
\nonumber\\&+B_{3}(3\cos^{2}\theta_{q}-1)(3\cos^{2}\theta_{p}-1)
+B_{4}\sin 2\theta_{p}\sin 2\theta_{q}
\cos\Delta\varphi\nonumber\\&
+B_{5}\sin^{2}\theta_{p}\sin^{2}\theta_{q}\cos 2\Delta\varphi\\
{\mathcal B}^{1}_{0}(1,1)&=C_{1}\sin 2\theta_{p}\sin 2\theta_{q}\sin\Delta\varphi
+C_{2}\sin^{2}\theta_{p}
\sin^{2}\theta_{q}\sin 2\Delta\varphi\\
{\mathcal B}^{2}_{0}(1,1)&=D_{0}+D_{1}(3\cos^{2}\theta_{q}-1)+D_{2}(3\cos^{2}
\theta_{p}-1)\nonumber\\&+D_{3}
(3\cos^{2}\theta_{q}-1)(3\cos^{2}\theta_{p}-1)+D_{4}\sin 2\theta_{p}\sin 2\theta_{q}
\cos\Delta\varphi\nonumber\\&
+D_{5}\sin^{2}\theta_{p}\sin^{2}\theta_{q}\cos 2\Delta\varphi.
\end{align}
\end{subequations}
The ${\mathcal B}^{k}_{q}(s_{i},s_{i}')$ may then be related to 
the $A_{ij}$'s defined in \cite{mey3} through
\begin{subequations}
\label{meyer-B}
\begin{eqnarray}
{\mathcal B}^{0}_{0}(0,0)&=& \sigma_{0}(\xi)\left[1-A_{\Sigma}(\xi)-A_{zz}
(\xi)\right]\\
{\mathcal B}^{0}_{0}(1,1)&=& \sigma_{0}(\xi)\left[3+A_{\Sigma}(\xi)+A_{zz}
(\xi)\right]\\
{\mathcal B}^{1}_{0}(1,1)&=& {\sqrt 6}\sigma_{0}(\xi)\left[A_{z0}(\xi)+A_{0z}
(\xi)\right]\\
{\mathcal B}^{2}_{0}(1,1)&=& 2{\sqrt 2}\sigma_{0}(\xi)  
\left[A_{zz}(\xi)-{\textstyle{\frac{1}{2}}}A_{\Sigma}(\xi)\right].
\end{eqnarray}
\end{subequations}
These relations enable us to identify the coefficients $A_{i},B_i,C_i,D_{i}$
in Eqs. (\ref{H}) in terms of the $E,F_{k},H_{k}^{ij},I,K$ of \cite{mey3}
for which numerical values have been deduced and given in 
table IV of \cite{mey3}.  It needs to be mentioned here that the 
$E,F_{k},H_{k}^{ij},I,K$ given in table
IV of \cite{mey3} are dimensionless as they have been normalized 
by a common factor of $(8\pi^2)/(\sigma_{tot})$, where
$\sigma_{tot}$ denotes the spin-averaged total cross section 
given in table V of \cite{mey3}.  As such, they
have to be multiplied
by $(\sigma_{tot})/(8\pi^2)$ before estimating the 
$A_{i},B_i,C_i,D_{i}$ of Eq. (\ref{H}) numerically and hence the
$^{2s_i +1}\sigma_{m}(\xi)$ of Eq. (\ref{ch-spin-cs}).
The one dimensional differential
cross sections are then defined as
\begin{subequations}
\begin{align}
{\frac{^{2s_{i}+1}d\sigma_{m}}{\sin\theta_q d\theta_q}}&=
\int {}^{2s_{i}+1}\sigma_{m}(\xi)\ d\Omega_p\ d\varphi_q\ d\epsilon\\
{\frac{^{2s_{i}+1}d\sigma_{m}}{\sin\theta_p d\theta_p}}&=
\int {}^{2s_{i}+1}\sigma_{m}(\xi)\ d\Omega_q\ d\varphi_p\ d\epsilon\\
{\frac{^{2s_{i}+1}d\sigma_{m}}{d\varphi_p d\varphi_q}}&=
\int {}^{2s_{i}+1}\sigma_{m}(\xi)\ d(\cos\theta_p)\ d(\cos\theta_q)
\ d\epsilon.
\end{align}   
\end{subequations}

\begin{center}
\begin{figure}[H]
\includegraphics[width=\linewidth]{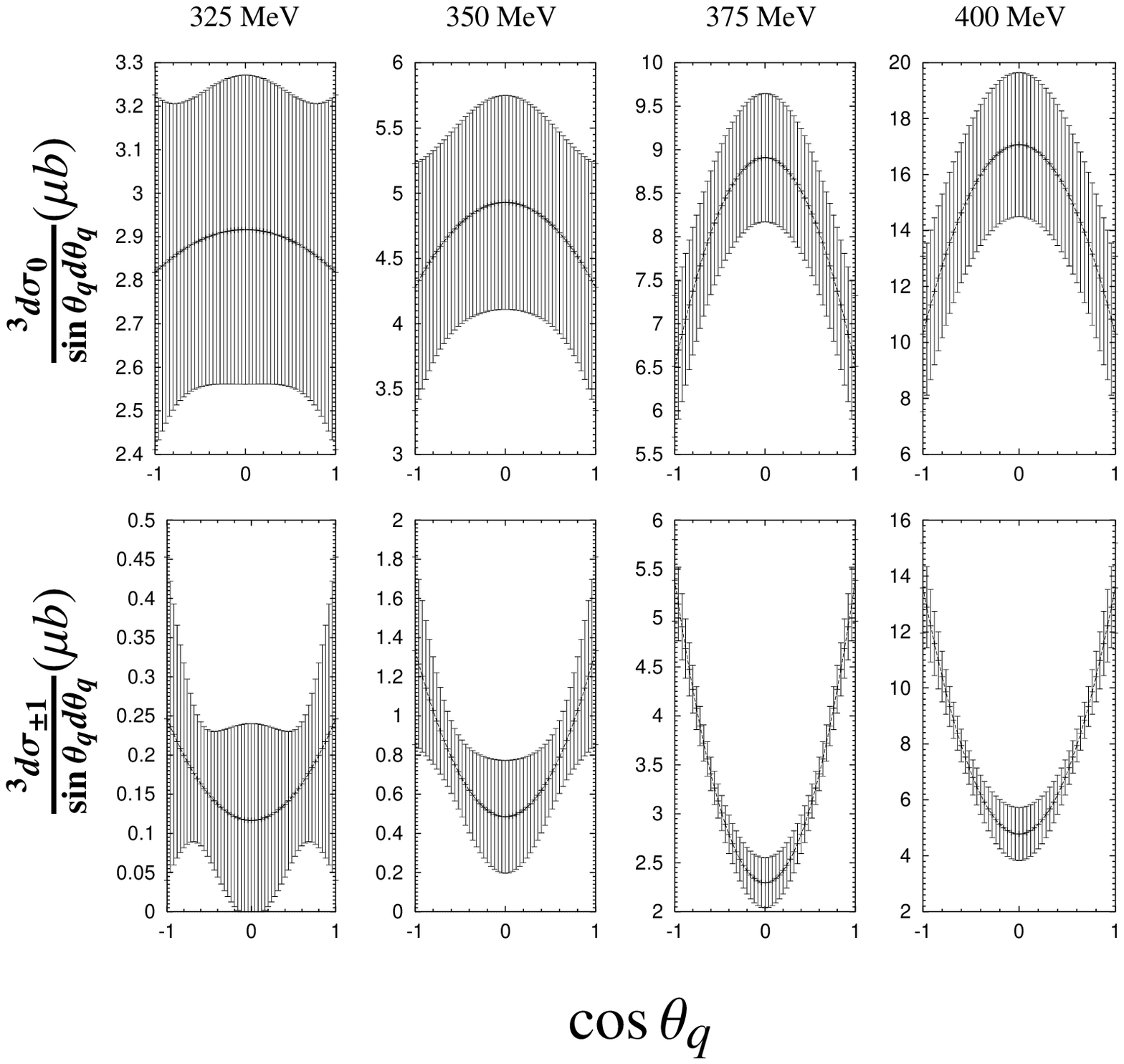}
\caption{A plot of $({}^{3}d\sigma_{m})/(\sin\theta_q d\theta_q),\ m=\pm 1,0$ as functions of
$\theta_q$ based on the results of Meyer et al \cite{mey3}.}
\label{fig1}
\end{figure}
\end{center}

\begin{center}
\begin{figure}[H]
\includegraphics[width=\linewidth]{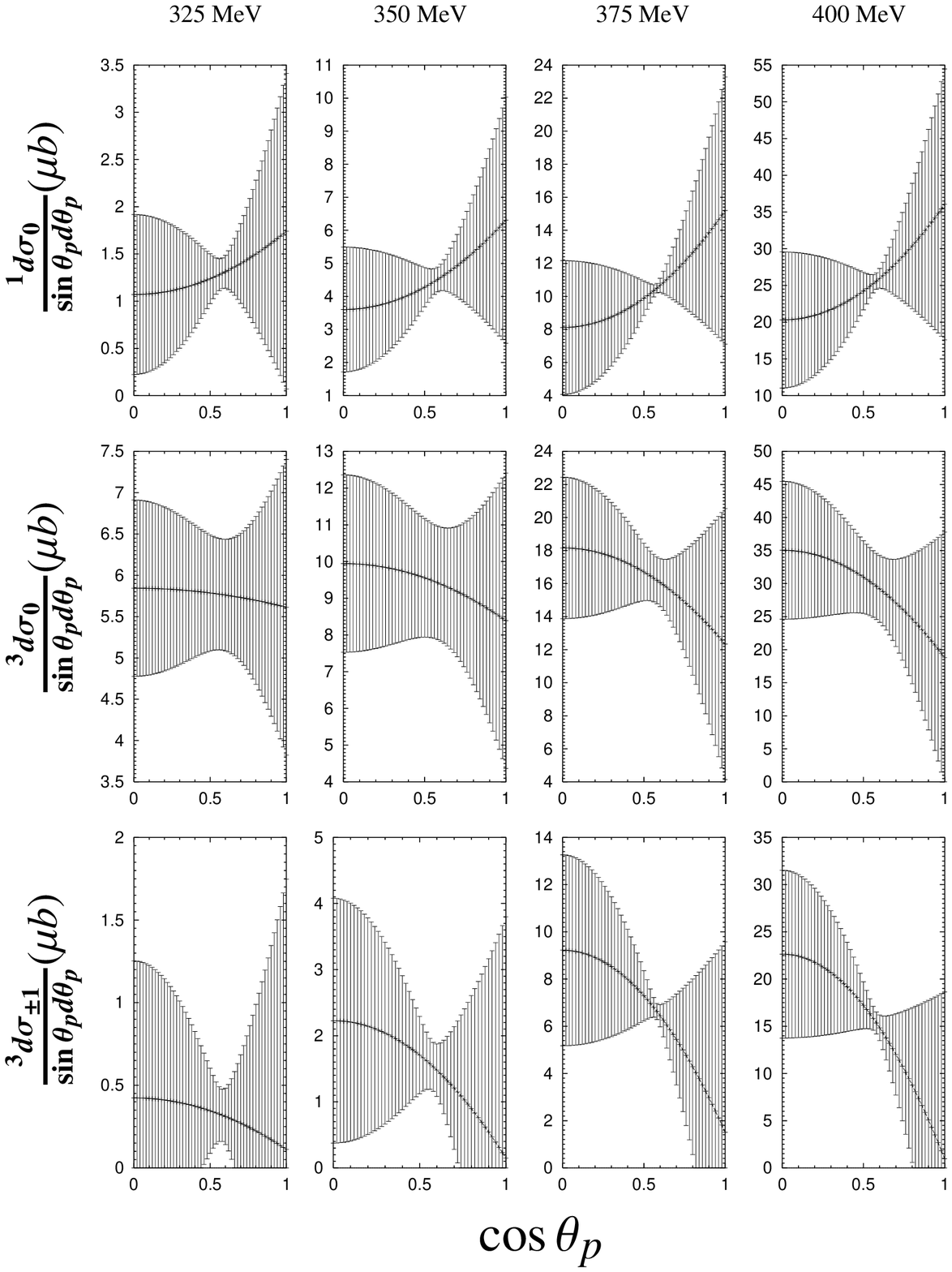}
\caption{A plot of $({}^{2s_i +1}d\sigma_{m})/(\sin\theta_p d\theta_p),\ m=\pm 1,0$ as functions of
$\theta_p$ based on the results of Meyer et al \cite{mey3}.}
\label{fig2}
\end{figure}
\end{center}

\begin{center}
\begin{figure}[H]
\includegraphics[scale=.85]{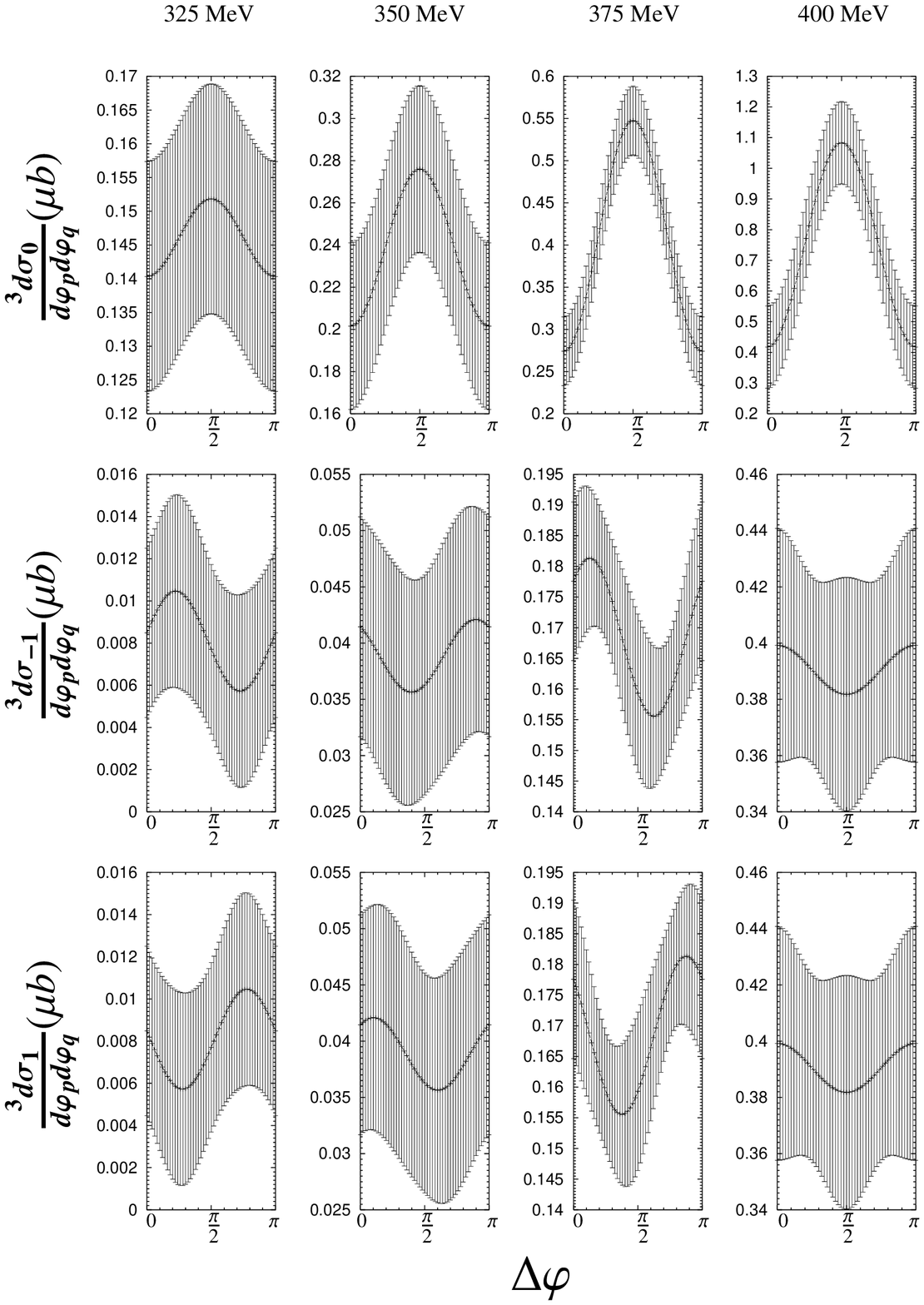}
\caption{A plot of $({}^{3}d\sigma_{m})/(d\varphi_p d\varphi_q),\ m=\pm 1,0$ as functions of
$\Delta\varphi$ based on the results of Meyer et al \cite{mey3}.}
\label{fig3}
\end{figure}
\end{center}

They may now be given explicitly in terms of the trigonometric functions of the respective
angles as 
\begin{subequations}
\label{trig}
\begin{eqnarray}
{\frac{^{3}d\sigma_{\pm 1}}{\sin\theta_q d\theta_q}}&=&
{\frac{\sigma_{tot}}{8}}\left[
\alpha_1 +\beta_1(3\cos^2\theta_q-1)
\right]\\
{\frac{^{3}d\sigma_{0}}{\sin\theta_q d\theta_q}}&=&
{\frac{\sigma_{tot}}{8}}\left[
\alpha_2 +\beta_2(3\cos^2\theta_q-1)
\right]\\
{\frac{^{1}d\sigma_{0}}{\sin\theta_q d\theta_q}}&=&
{\frac{\sigma_{tot}}{2}}F_1\\
{\frac{^{3}d\sigma_{\pm 1}}{\sin\theta_p d\theta_p}}&=&
{\frac{\sigma_{tot}}{4}}\left[
\alpha_1 +\beta_3(3\cos^2\theta_p-1)
\right]\\
{\frac{^{3}d\sigma_{0}}{\sin\theta_p d\theta_p}}&=&
{\frac{\sigma_{tot}}{4}}\left[
\alpha_2 +\beta_4(3\cos^2\theta_p-1)
\right]\\
{\frac{^{1}d\sigma_{0}}{\sin\theta_p d\theta_p}}&=&
\sigma_{tot}\left[F_1 + \beta_5(3\cos^2\theta_p-1)
\right]\\
{\frac{^{3}d\sigma_{\pm 1}}{d\varphi_p d\varphi_q}}&=&
{\frac{\sigma_{tot}}{8\pi^2}}
\left[
\alpha_3 + \beta_6 \cos 2\Delta\varphi\pm\beta_7\sin 2\Delta\varphi
\right]\\
{\frac{^{3}d\sigma_{0}}{d\varphi_p d\varphi_q}}&=&
{\frac{\sigma_{tot}}{8\pi^2}}\left[
\alpha_4 + \beta_8 \cos 2\Delta\varphi
\right]\\
{\frac{^{1}d\sigma_{0}}{d\varphi_p d\varphi_q}}&=&
{\frac{\sigma_{tot}}{4\pi^2}}F_1,
\end{eqnarray}
\end{subequations} 
where the coefficients
$\alpha_i,\ \beta_i$ are once again dimensionless since
$\sigma_{tot}$ occurs as a common factor in each of the Eqs. (\ref{trig}). 
It may be noted that $F_1$ is the same as in \cite{mey3}.
The numerical values of all these coeffecients together with their
errors are presented in Table 1.
\begin{table}[H]
{\footnotesize{ 
\caption{Values of the coefficients occurring in Eqs. (\ref{trig})
at the four bombarding energies  
based on the results of Meyer et al \cite{mey3}.}
\begin{tabular}{cccccccccccccccc}\hline\hline
\hspace*{.2cm}&\multicolumn{3}{c}{$325$ MeV}&&\multicolumn{3}{c}{$350$ MeV}&&
\multicolumn{3}{c}{$375$ MeV}&&\multicolumn{3}{c}{$400$ MeV}\\
&Value&&Error&&Value&&Error&&Value&&Error&&Value&&Error\\\hline
$\alpha_{1}$&$0.166$&&$0.082$&&$0.361$&&$0.087$&&$0.665$&&$0.034$&&$0.717$&&$0.054$\\
$\alpha_{2}$&$2.996$&&$0.349$&&$2.218$&&$0.364$&&$1.622$&&$0.142$&&$1.378$&&$0.233$\\
$\alpha_{3}$&$0.083$&&$0.041$&&$0.181$&&$0.043$&&$0.333$&&$0.017$&&$0.359$&&$0.027$\\
$\alpha_{4}$&$1.498$&&$0.174$&&$1.109$&&$0.182$&&$0.811$&&$0.071$&&$0.689$&&$0.116$\\
$\beta_{1}$&$0.045$&&$0.099$&&$0.133$&&$0.104$&&$0.206$&&$0.038$&&$0.273$&&$0.07$\\
$\beta_{2}$&$-0.034$&&$0.12$&&$-0.102$&&$0.128$&&$-0.16$&&$0.04$&&$-0.21$&&$0.056$\\
$\beta_{3}$&$-0.054$&&$0.423$&&$-0.163$&&$0.427$&&$-0.257$&&$0.403$&&$-0.336$&&$0.41$\\
$\beta_{4}$&$-0.04$&&$0.431$&&$-0.122$&&$0.437$&&$-0.194$&&$0.404$&&$-0.252$&&$0.426$\\
$\beta_{5}$&$0.029$&&$0.108$&&$0.053$&&$0.109$&&$0.059$&&$0.101$&&$0.061$&&$0.107$\\
$\beta_{6}$&$0.004$&&$0.004$&&$0.012$&&$0.013$&&$0.018$&&$0.019$&&$0.008$&&$0.027$\\
$\beta_{7}$&$-0.024$&&$0.023$&&$0.009$&&$0.021$&&$-0.018$&&$0.009$&&$0$&&$0.014$\\
$\beta_{8}$&$-0.059$&&$0.008$&&$-0.173$&&$0.024$&&$-0.270$&&$0.038$&&$-0.306$&&$0.04$\\
$F_{1}$&$0.168$&&$0.021$&&$0.265$&&$0.022$&&$0.262$&&$0.007$&&$0.297$&&$0.013$\\\hline\hline
\end{tabular}
\label{tab1}
}}
\end{table}
Of the total 16 partial wave amplitudes taken into consideration, it 
is clear that 11 amplitudes : 
$\mathsf{M}^{0}_{1(11)1;11},\ \mathsf{M}^{1}_{1(11)0;11},
\ \mathsf{M}^{1}_{1(11)1;11},\ \mathsf{M}^{1}_{1(11)2;11}$,
$\mathsf{M}^{2}_{1(11)1;11},\ \mathsf{M}^{2}_{1(11)2;11},
\ \mathsf{M}^{2}_{1(11)1;31},$\newline$\mathsf{M}^{2}_{1(11)2;31},
\ \mathsf{M}^{3}_{1(11)2;31},\mathsf{M}^{0}_{0(11)0;00},\ \mathsf{M}^{2}_{0(11)2;20}$ 
lead to $s_f =1$ and 
5 amplitudes : $\mathsf{M}^{0}_{0(00)0;11},$\newline
$\mathsf{M}^{2}_{0(20)2;11},
\ \mathsf{M}^{2}_{0(20)2;31},\ \mathsf{M}^{2}_{2(00)0;11},
\ \mathsf{M}^{2}_{2(00)0;31}$ 
lead to $s_{f}=0$.  It is also clear
from Eq. (\ref{bilinears}) that these two sets do not mutually interfere.
This has been noted also in \cite{mey3}.  It may also be noticed
that only two among the 16 viz., $\mathsf{M}^{0}_{0(11)0;00}$
and $\mathsf{M}^{2}_{0(11)2;20}$ are the singlet amplitudes.  Both
of them lead to $l_q =0$.  Hence it follows that the singlet 
differential cross section 
$({}^{1}d\sigma_{0})/(\sin\theta_q d\theta_q)$
 is independent 
of $\theta_q$ which is clear from Eqs. (\ref{ch-spin-1}) and
(\ref{b-singlet}).  From these two equations it can also be seen that
$({}^{1}d\sigma_{0})/(d\varphi_p d\varphi_q)$ is independent of $\Delta\varphi$.
However, the singlet differential cross section
$({}^{1}d\sigma_{0})/(\sin\theta_p d\theta_p)$
varies with $\theta_p$ since $l_p = 1$
in either of the singlet partial wave amplitudes.  The dependence
on the angle $\theta_p$ is explicitly seen from Eqs. (\ref{ch-spin-1}) and 
Eqs. (\ref{b-singlet}).   The estimates deduced for the triplet differential cross sections
$({}^{3}d\sigma_{m})/(\sin\theta_q d\theta_q)$ are given in Figure \ref{fig1}, while 
$({}^{1}d\sigma_{0})/(\sin\theta_p d\theta_p)$ and 
$({}^{3}d\sigma_{m})/(\sin\theta_p d\theta_p)$
are shown in Figure \ref{fig2}, whereas 
$({}^{3}d\sigma_{m})/(d\varphi_p d\varphi_q)$ as a function of
$\Delta\varphi$ are shown in Figure \ref{fig3} for the four bombarding 
energies 325, 350, 375, 400 MeV.
The estimates for $({}^{1}d\sigma_{0})/(\sin\theta_q d\theta_q)$ 
and $({}^{1}d\sigma_{0})/(d\varphi_p d\varphi_q)$
which are independent of
$\theta_q$ and $\Delta\varphi$ respectively are given in 
Table 2 for the bombarding energies 325, 350, 375 and 400 MeV.
It may be noted that $\mathcal{B}^1_0(1,1)$
vanishes on integration with respect to $d\varphi_q\ d\varphi_p$.  Therefore
$({}^{3}d\sigma_{1})/(\sin\theta_q d\theta_q)=({}^{3}d\sigma_{-1})
/(\sin\theta_q d\theta_q),({}^{3}d\sigma_{1})/(\sin\theta_p d\theta_p)=
({}^{3}d\sigma_{-1})/
(\sin\theta_p d\theta_p)$ whereas $({}^{3}d\sigma_{1})/(d\varphi_p d\varphi_q)\neq
({}^{3}d\sigma_{-1})/(d\varphi_p d\varphi_q)$. 
\begin{table}[H]
\caption{Numerical estimates deduced for the (angle-independent)
$({}^{1}d\sigma_{0})/(\sin\theta_q d\theta_q)$ and
$({}^{1}d\sigma_{0})/(d\varphi_p d\varphi_q)$
at the four bombarding
energies based on the results of Meyer et al \cite{mey3}.}
\begin{tabular}{ccccccccc}\hline\hline
&\multicolumn{2}{c}{$325$ MeV}&\multicolumn{2}{c}{$350$ MeV}&\multicolumn{2}{c}{$375$ MeV}
&\multicolumn{2}{c}{$400$ MeV}\\
& Value & Error & Value & Error & Value & Error & Value & Error\\
&\multicolumn{2}{c}{($\mu$b)}&\multicolumn{2}{c}{($\mu$b)}&\multicolumn{2}{c}{($\mu$b)}
&\multicolumn{2}{c}{($\mu$b)}\\\hline
&&&&&&&\\
$(^{1}d\sigma_{0})/(\sin\theta_q d\theta_q)$&$0.647$&$0.081$&$2.253$&$0.187$
&$5.24$&$0.140$&$12.771$&$0.559$\\
&&&&&&&&\\
$(^{1}d\sigma_{0})/(d\varphi_p d\varphi_q)$&$0.033$&$0.004$&$0.114$&$0.010$
&$0.266$&$0.007$&$0.647$&$0.028$\\
&&&&&&&\\\hline\hline
\end{tabular}
\end{table}

The numerical values for $E,F_{k},H^{ij}_k,I,K$ given in table IV of
\cite{mey3} together with $\sigma_{tot}$ given in table V of
\cite{mey3} are adequate to deduce the above estimates for the
differential cross sections $^{2s_{i}+1}\sigma_{m}$ at the one 
dimensional level.  It may, however, be noted from Eqs. (\ref{ch-spin-cs}),
(\ref{H}) and (\ref{meyer-B}) that one needs also estimates (based
on experiments) for $H^{00}_{3},H_3^{\Sigma},H_{3}^{zz}$
in addition to those given in table IV of \cite{mey3} in order
to estimate $^{2s_{i}+1}\sigma_{m}(\xi)$ at the double
differential level.  It would therefore be desirable to experimentally
determine the $H^{00}_{3},H_3^{\Sigma},H_{3}^{zz}$.  We
may point out further that the double differential cross sections
$^{2s_{i}+1}\sigma_{m}(\xi)$ may also be directly determined from
experimental measurements as suggested in \cite{gr1}.
Therefore we would like to encourage further measurements of the 
double differential cross sections for $\vec{p}\vec{p}\to pp\pi^\circ$
on the above lines.

We thank H. O. Meyer for making \cite{mey3} available before publication and
for very useful correspondence.
We thank the Council of Scientific and Industrial Research (CSIR), India 
for support.

\end{document}